\documentclass[11pt,twoside]{article}

\usepackage{asp2006}
\usepackage{lscape}

\defcitealias{cww80}{Coleman et al.(1980)}

\begin{document}

\title{Low Resolution Spectral Templates for AGNs and Galaxies from
0.03 -- 30 $\mu$m} %%% Fill in title

\author{R.J.~Assef,
  C.S.~Kochanek \& The AGES Collaboration} 

\affil{Department of Astronomy, The Ohio State University,\\ 140 W.\
18th Ave., Columbus, OH 43210\\e-mail: {\tt{rjassef@astronomy.ohio-state.edu}}}

\begin{abstract} %%% Abstract to run on from here.
We present a set of low resolution empirical SED templates for AGNs
and galaxies in the wavelength range from 0.03 to 30$\mu$m. These
templates form a non-negative basis of the color space of such objects
and have been derived from a combination 14448 galaxies and 5347
likely AGNs in the NDWFS Bo\"otes field. We briefly describe how the
templates are derived and discuss some applications of them. In
particular, we discuss biases in commonly used AGN mid-IR color
selection criteria and the expected distribution of sources in the
current WISE satellite mission.
\end{abstract}

%%% MAIN BODY OF TEXT GOES HERE. CONSULT "INSTRUCTIONS FOR AUTHORS USING
%%% LATEX2E MARKUP", SECTIONS 2.3-2.6 FOR HELP WITH EQUATIONS, FIGURES,
%%% AND TABLES.

%\section{}   %%% Top level section head (remove "%" symbol)
%\subsection{}   %%% Second level section head (remove "%" symbol)
%\subsubsection{}   %%% Lowest level section head (remove "%" symbol)
%\section*{}    %%% Unnumbered top level section head (remove "%" symbol)
%\subsection*{}   %%% Unnumbered second level section head (remove "%" symbol)

\section{Introduction}

In many current and upcoming extragalactic photometric surveys it will
be necessary to use spectral energy distribution (SED) fitting
techniques to characterize most objects (e.g., redshifts,
K-corrections, stellar masses and bolometric luminosities), because
spectroscopic observations are too expensive.

SEDs of galaxies are typically either empirically obtained
\citep[e.g.,][]{cww80} or theoretically created
\citep[e.g.,][]{bc03}. The former have the advantage of giving an
accurate representation of the SEDs but generally lack the large
wavelength ranges of theoretical ones. Empirical SEDs also have the
advantage of easily including non-stellar emission, like active nuclei
and dust/PAH emission, which are hard to model from first
principles. In particular, most available AGN SEDs are empirically
obtained \citep[e.g.,][]{richards06} by averaging photometric
observations of a small number of quasars.

In this proceeding we discuss a non-negative basis of empirically
obtained SED templates for galaxies and AGNs that accurately
represents the colors of such objects. These templates are derived
from a combination of 14448 galaxies and 5347 likely AGNs in the NDWFS
Bo\"otes field with spectroscopic redshifts and photometry spanning
the far-UV to the mid-IR. All results shown in here have been
presented and are discussed in detail by \citet{assef08},
\citet{assef10a} and \citet{assef10b}.

\section{Data}\label{sec:data}

Our data set consists on 14448 galaxies and 5347 likely AGNs in the
NOAO Deep Wide-Field Survey \citep[NDWFS;][]{ndwfs99} Bo\"otes
field. Each object has a spectroscopic redshift from the AGN and
Galaxy Evolution Survey \citetext{AGES; Kochanek et al. in
preparation} and has at least an upper limit in 8 of the 14 following
bands: Bw, R, I and K from NDWFS, J and Ks from FLAMEX, z from
zBo\"otes, IRAC [3.6], [4.5], [5.8] and [8.0] from SDWFS, GALEX FUV
and NUV, and MIPS 24$\mu$m from \citet{weedman06}. The AGN
classification scheme is conservative, and objects classified as AGNs
are either spectroscopically or photometrically classified according
to their mid-IR, X-ray and radio properties. We refer the reader to
\citet{assef10a} for details on the sample selection.

\section{Building the SED Templates}\label{ssec:met_temps}

The methods we use to derive the SED templates are described in detail
by \citet{assef08} and \citet{assef10a}. In this section we give a
brief summary of how our methods work and refer the reader to the
aforementioned papers for detailed discussions.

We assume that the spectrum of any object in our sample can be modeled
as a linear combination of a small set of unknown spectral
templates. We construct these unknown templates using the data sets
described in the previous section. For ``pure'' galaxies, we assume
the majority of them have SEDs that can be described as a linear
combination of three templates: one similar to an elliptical galaxy
(an old stellar population), one similar to a spiral galaxy (a
continuously star forming population), and a third similar to an
irregular galaxy (a starburst population). For objects with active
nuclei, we assume that every AGN SED can be described by the same
spectral template with varying amounts of reddening and absorption by
the intergalactic medium (IGM), combined with the galaxy templates to
describe the host. 

We use an iterative algorithm to derive the SED templates. This
algorithm is based on the method proposed by \citet{budavari00} for
deriving principal component SEDs. We start with an initial guess for
each of our SED templates and sequentially improve it to best fit the
photometric data. Our initial guess for the ``pure'' galaxy SEDs
corresponds to the E, Sbc and Im templates of \citetalias{cww80},
extended into the UV and mid-IR with the models of \citet{bc03} and
with the addition of PAH and dust emission of the VCC1003 and M82 SEDs
obtained by \citet{devriendt99} to the star forming templates. For the
AGN component we use a combination of power-laws that broadly
resembles the AGN SED template of \citet{richards06}. We divide every
SED template in 300 logarithmically spaced wavelength bins and fit
each bin independently to our data set, with a restriction to force
their smoothness. The top left panel of Figure \ref{fg:temps} shows
the resulting SED templates alongside with the initial guesses.

\section{Applications}

We have pursued several applications of our SED templates, including
the estimation of photometric redshifts and K-corrections for galaxies
and AGNs, the study of biases in mid-IR selection of AGNs, predictions
of colors of AGNs and galaxies in the pass-bands of the recently
launched WISE mission, and the determination of luminosity function of
mid-IR and X-ray selected AGNs from redshift 0 to 6. The former three
applications are discussed in detail by \citet{assef08} and
\citet{assef10a}, while our results on the AGN luminosity function
were presented by \citet{assef10b}. In this section we discuss two of
these applications, namely the biases in the mid-IR color selection of
AGNs and the colors of AGNs and galaxies in WISE.

\begin{figure}[!ht]
  \begin{center}
    \plottwo{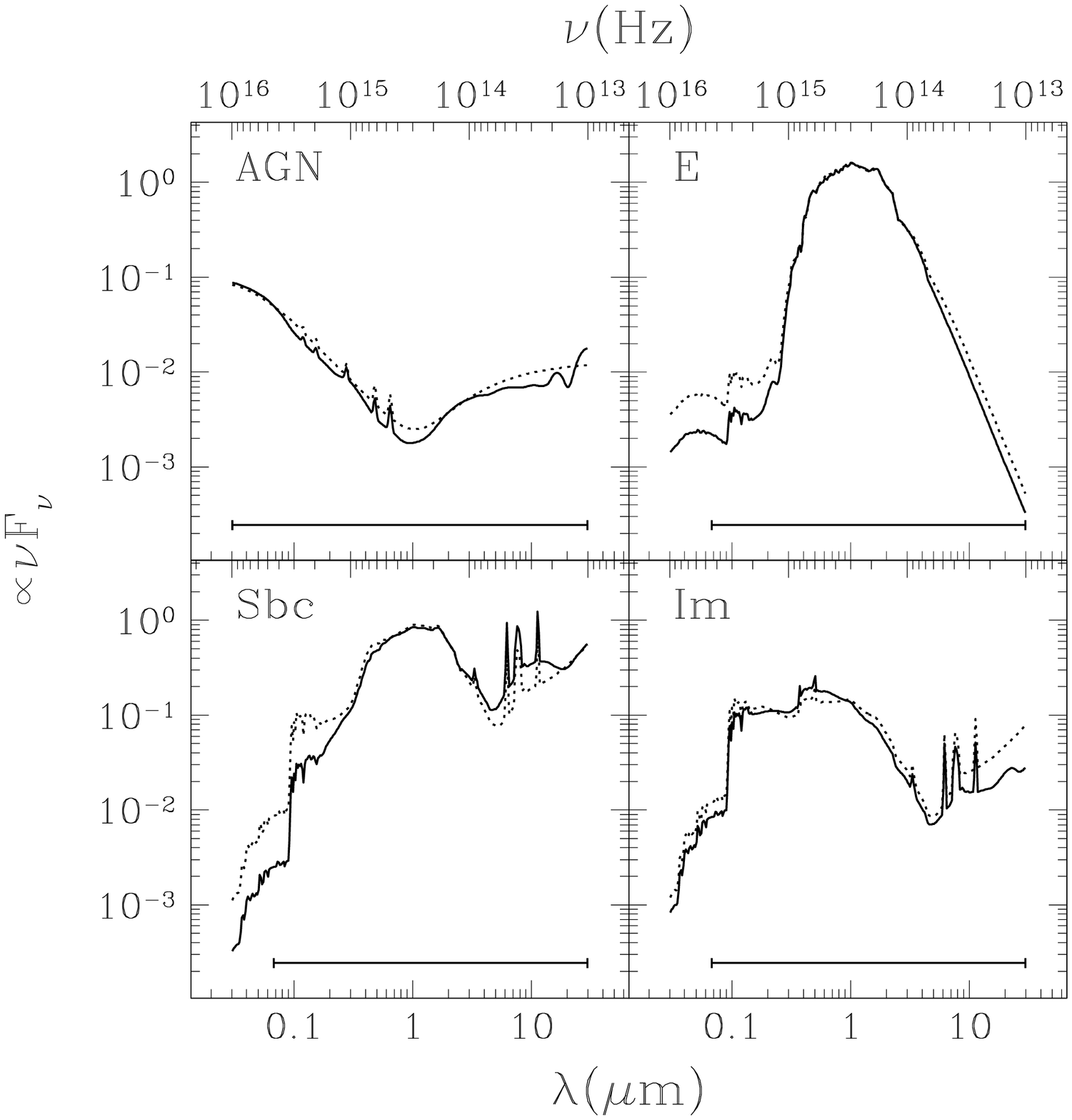}{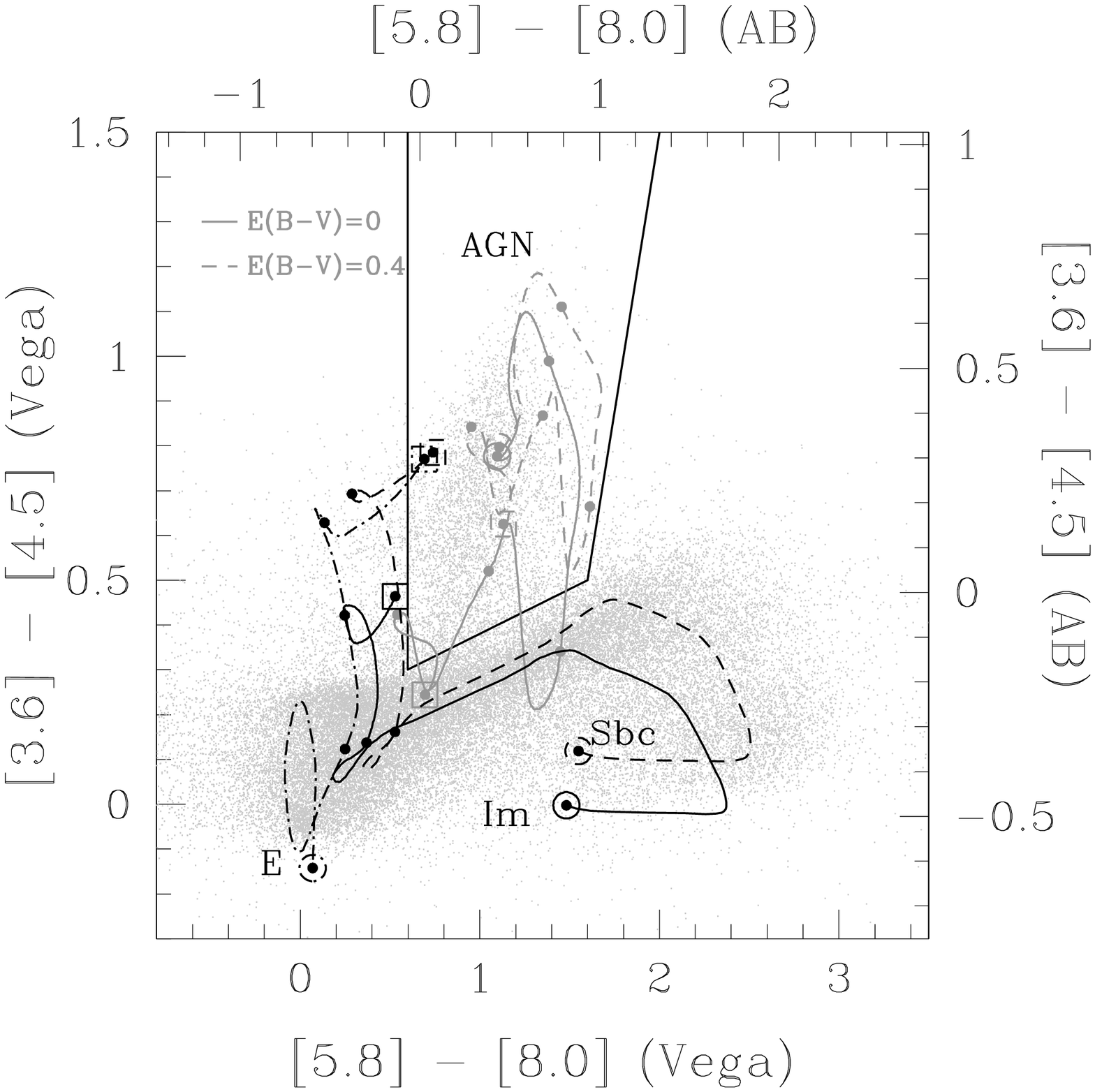}
    \plottwo{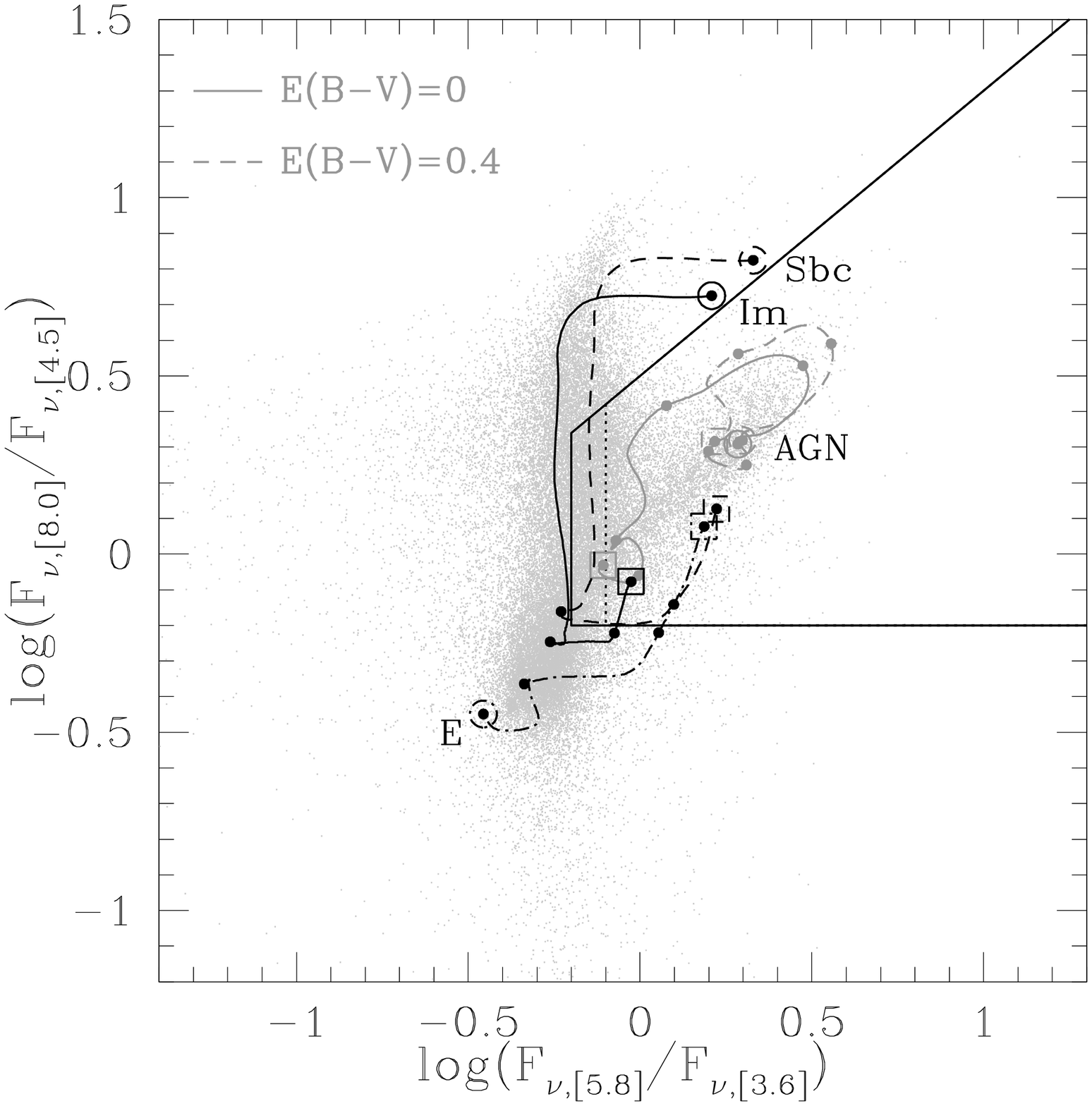}{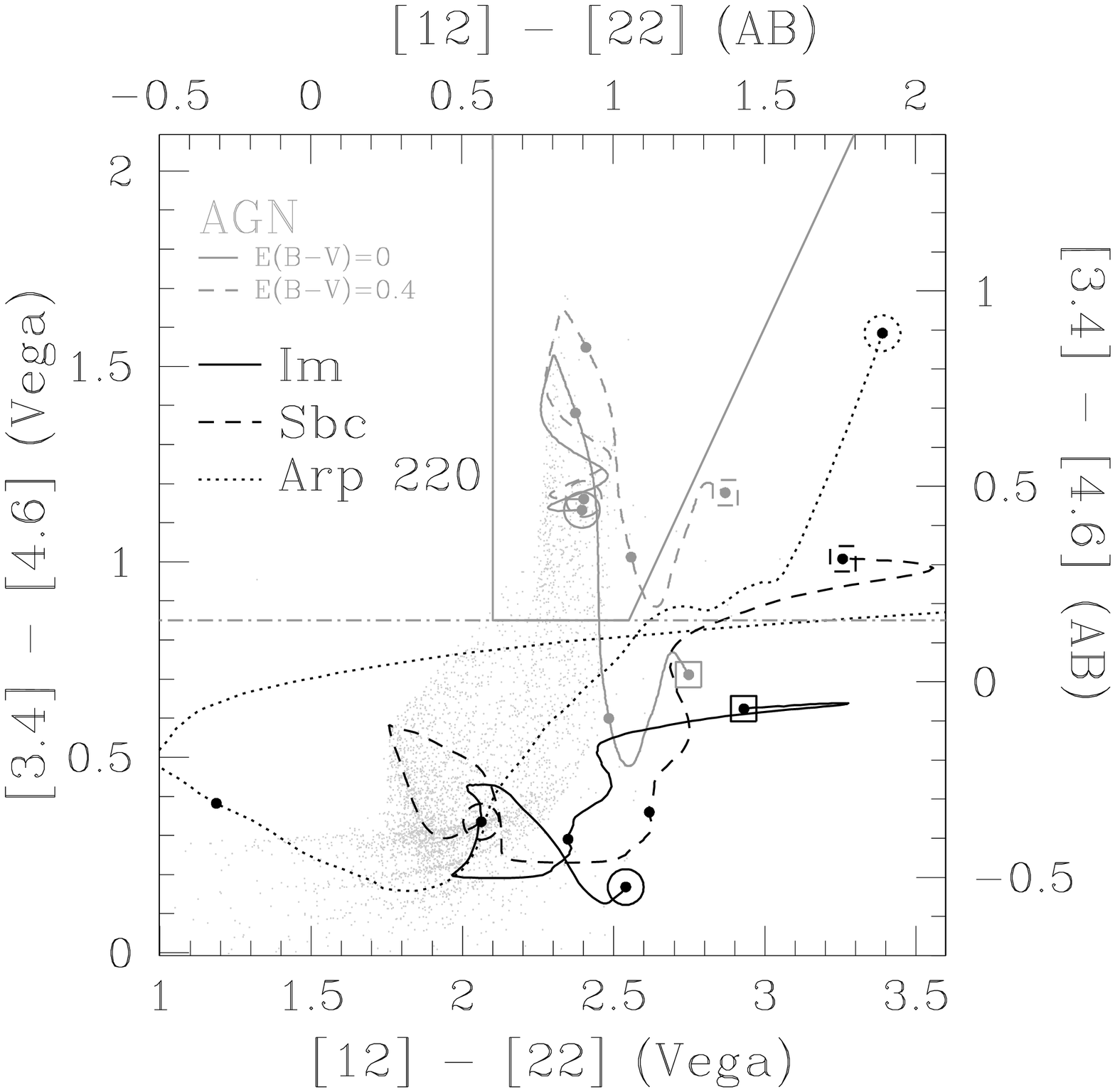}
    \caption{({\it{Top Left}}) Resulting best fit templates from
      applying the algorithm of \S\ref{ssec:met_temps} to the data set
      described in \S\ref{sec:data} ({\it{black solid}}) compared to
      their initial guesses ({\it{black dotted}}). The bars at the
      bottom of each panel show the rest-frame wavelength coverage of
      our data set. ({\it{Top Right}}) IRAC band color-color diagram
      of $I\leq 21.5$ SDWFS sources ({\it{gray dots}}). Overlaid are
      the AGN selection region of \citet{stern05} ({\it{solid black
          boundaries}}) and the color tracks (as a function of
      redshift) of our galaxy SED templates and of our AGN template
      without reddening and with $E(B-V) = 0.4$. For the galaxy
      templates, color tracks are shown for redshifts between 0--3,
      while for the AGN template they are shown between 0--10. Each
      heavy dot marks an increase of unity in redshift for the
      galaxies and an increase of 2 units of redshift for the AGN. The
      heavy dots surrounded by a circle marks $z=0$ while those
      surrounded by a square mark the terminal redshift. ({\it{Bottom
          Left}}) Same as top right panel but in the colors used by
      \citet{lacy04} to define their AGN selection criterion
      ({\it{solid gray boundaries}}). The dotted line shows a revised
      version of this criteria by \citet{lacy07}. ({\it{Bottom
          Right}}) WISE color tracks of our SED templates and of the
      SED of Arp 220, overlaid on top of all SDWFS sources that would
      be detected at [3.3] and [4.6] by WISE ({\it{gray dots}}). The
      different line-styles have the same definition as in the top
      right panel. Note that, because of the shallow depths of the
      WISE mission, we only show the color tracks up to $z=2$, 6 and 3
      for galaxies, AGN and Arp 220 respectively. Most of the Arp 220
      track extends off the red edge of the Figure while the entire E
      color track is to the blue.}
    \label{fg:temps}
    \epsscale{1}
  \end{center}
\end{figure}

Figure \ref{fg:temps} shows the mid-IR colors of all sources in SDWFS
with $I\leq 21.5$, the AGN selection diagrams of \citet{stern05} (top
right panel) and of \citet{lacy04} (bottom left panel) and the color
tracks of our templates. The boundaries of the selection region of
\citet{stern05} do a remarkable job of excluding inactive galaxies at
$z<3$, but also exclude unreddened AGNs at $z\sim 4.5$. In a more
subtle manner, the selection region of \citet{stern05} also excludes
AGNs that are faint in comparison to their host galaxies, as the
latter will dominate the mid-IR colors. The selection region proposed
by \citet{lacy04} is not strongly affected by the biases of the
\citet{stern05} criteria, but their AGN selection region is heavily
contaminated by intermediate redshift star forming galaxies and
galaxies at $z>2$. These problems are partially solved by the revised
version of this criteria by \citet{lacy07}. It can be seen, however,
that if the left-most boundary in this criterion is pushed further to
the right to limit contamination by inactive galaxies, similar biases
to those of the \citet{stern05} selection criteria arise.

Using our SED templates, we have also studied the color distribution
of sources in the WISE mission. The selection boundaries shown in the
bottom right panel of Figure \ref{fg:temps} correspond to those
proposed by \citet{assef10a}, and would be able to classify in between
$\sim 750,000$ and $\sim 1,500,000$ AGNs, depending on whether a
detection in the two longest wavelength bands is required or not. This
criterion would keep the contamination by inactive galaxies and ULIRGS
to a minimum, as shown by the color tracks of our SEDs and of the SED
of Arp 220 produced by GRASIL \citep{silva98}. We note, however, that
the WISE colors of $z\ga 3$ AGNs are similar to those of intermediate
redshift galaxies and ULIRGS, leading to a color degeneracy that must
be broken by the inclusion of other data (e.g., optical broad-band
photometry).

%\acknowledgements %%% Text of acknowledgements runs on after this command.

%%% THE BIBLIOGRAPHY
%%%
%%% CONSULT SECTION 3 OF "INSTRUCTIONS FOR AUTHORS" FOR HOW TO USE NATBIB.
%%% AUTHORS ARE ENCOURAGED TO USE EITHER THE "THEBIBLIOGRAPY" ENVIRONMENT
%%% BY UNCOMMENTING (DELETING THE "%" SYMBOL) THE COMMANDS BELOW, OR BY
%%% USING THE BIBTEX ENVIRONMENT. TO FIND OUT WHICH IS APPLICABLE TO YOUR
%%% CONTRIBUTION, CONSULT THE VOLUME EDITORS FOR YOUR PROCEEDINGS.
%%%

%\begin{thebibliography}{}
%\bibitem[]{}
%\bibitem[]{}
%\bibitem[]{}
%\bibitem[]{}
%\bibitem[]{}
%\bibitem[]{}
%\bibitem[]{}
%\bibitem[]{}
%\bibitem[]{}
%\bibitem[]{}
%\bibitem[]{}
%\bibitem[]{}
%\end{thebibliography}

\end{document}